\documentclass[12pt]{article}

\input epsf
\usepackage{amssymb,wrapfig}
\usepackage{xypic,multirow}
\usepackage[matrix,arrow,curve]{xy}

\begin{document}
\addtolength{\baselineskip}{.5mm}
\newlength{\extraspace}
\setlength{\extraspace}{1.5mm}
\newlength{\extraspaces}
\setlength{\extraspaces}{2mm}

\def\nn{{\cal N}}
\def\rr {{\Bbb R}}
\def\cc {{\Bbb C}}
\def\pp {{\Bbb P}}
\def\zz {{\Bbb Z}}
\def\del {\partial}
\def\cy {Calabi--Yau}
\def\ka {K\"ahler}
\newcommand{\inv}[1]{{#1}^{-1}} 

\newcommand{\Section}[1]{\section{#1} \setcounter{equation}{0}}

\makeatletter
\@addtoreset{equation}{section}
\makeatother
\renewcommand{\theequation}{\thesection.\arabic{equation}}

\begin{center}

{\hfill SLAC--PUB--12049\\ \hfill SU--ITP--06/21}

{\Large \bf{Topological twisted sigma model with H-flux
revisited}}

\bigskip

Wu-yen Chuang

\medskip

{\it ITP, Stanford University, Stanford, CA 94305, USA}

{\it SLAC, Stanford University, Menlo Park, CA 94025, USA }

{\it wychuang@stanford.edu}

\bigskip
{\bf Abstract}
\end{center}
In this paper we revisit the topological twisted sigma model with
H-flux. We explicitly expand and then twist the worldsheet
Lagrangian for bi-Hermitian geometry. We show that the
resulting action consists of a BRST exact term and  pullback
terms, which only depend on one of the two generalized complex
structures and the B-field. We then discuss the topological feature
of the model.

\section{Introduction}

It is a very convenient and powerful approach to obtain topological field theories by
twisting supersymmetric field theory \cite{Witten1}. It was furthur shown that the $N=(2,2)$
worldsheet sigma model with the K\"ahler target space admits A and B types of twisting \cite{Witten}.
However the K\"ahler condition is not crucial to perform the A and B twists. What is really needed is
to have $N=(2,2)$ worldsheet supersymmetry so that $U(1)_V$ and $U(1)_A$ exist.

From the viewpoint of the $N=(2,2)$  worldsheet supersymmetry algebra the twists are achieved by
replacing the 2d Euclideanized spacetime rotation group $U(1)_E$ with the diagonal
subgroup of $U(1)_E \times U(1)_R$, where $U(1)_R$ is either $U(1)_V$ or $U(1)_A$ R-symmetry
in the $N=(2,2)$ supersymmetry group.

In 1984 the most general geometric backgrounds for $N=(2,2)$
supersymmetric sigma models was proposed by Gates, Hull, and
Ro$\check{c}$ek \cite{GHR}. The geometric backgrounds (a.k.a.
bi-Hermitian geometry) consists of a set of data $(J_{+}, J_{-}, g,
H)$. $J_{\pm}$ are two different integrable complex structures and
the metric $g$ is Hermitian with respect to either one of $J_{\pm}$.
Moreover $J_{\pm}$ are covariantly constant with respect to the
torsional connections $\Gamma \pm g^{-1} H$, where $H$ is a closed
3-form on the manifold. The manifold is apparently non-K\"ahler due
to the presence of the torsions.

Bi-Hermitian geometry started to re-receive new attention after Hitchin introduced the notion of generalized geometry \cite{Hitchin}
and Gualtieri furthur showed that the geometry is equivalent to a pair of commuting (twisted) generalized complex  ((T)GC for short) structures on the manifold $M$, namely, the twisted generalized K\"ahler structure \cite{Gual}.

Since the worldsheet theory with bi-Hermitian target has $N=(2,2)$
supersymmetry, we definitely can consider its topological twisted
models. In \cite{AK} Kapustin and Li considered such a topological
model and showed that on the classical level the topological
observables in a given twisted model correspond to the Lie algebroid
cohomologies associated with one of the two twisted generalized
complex structures. The same problem was also considered by many
other authors from Hamiltonian approach or using Batalin-Vilkovisky
quantization \cite{Maxim1} \cite{Maxim2} \cite{Zucch1}.

Although it is definitely true that the twisted models for
bi-Hermitian geometries are topological, the explicit construction
of the twisted Lagrangian is lacking. The difficulties of such a
calculation lie in that people are so accustomed to using complex
geometry that they feel reluctant to perform a calculation which
needs to be done in the real coordinate basis with projectors. A
priori, we should be able to express the twisted Lagrangian for the
generalized geometry as some BRST exact piece plus certain pullback
terms which only depend on one of the twisted generalized complex
structures.

By the end of the paper we will see that this is indeed true.
However since the pullback object is not closed it is not clear that
the action is topological. This issue is made clear in
\cite{Zucch2}. The paper is organized as follows. In Section
\ref{review} we first review the sigma models with Riemannian and
K\"ahler targets and discuss the properties of the twisted
Lagrangian. In Section \ref{biH} we present the computation of the
twisted topological models for bi-Hermitian geometries and express
the twisted Lagrangian in the aforementioned way. In section
\ref{con} we conclude, discuss the limitation of the twisted models,
and mention some open questions. The equivalence between on-shell
and off-shell $(2,2)$ worldsheet supersymmetry and some basics of
the generalized geometry will be presented in the appendix.

\section{Topological sigma model with K\"ahler targets}
\label{review}

We first recall some basic facts about the worldsheet sigma models
with Riemannian or K\"ahler manifolds as targets. Throughout the
whole paper lowercase English letters $a,b,c,...$ are indices for
the real coordinates on the targets, while Greek letter $\mu, \nu,
\sigma,...$ are those for holomorphic coordinates. (And of course
$\bar{\mu}, \bar{\nu}, \bar{\sigma},...$ for antiholomorphic
coordinates.)

The nonlinear sigma model with a Riemannian manifold $M$ has
natural $(1,1)$ worldsheet supersymmetric formalism. The model is
governed by an embedding map $\Phi : \Sigma \to M$ where $\Sigma$
is a Riemann surface. The Lagrangian is

\begin{equation}
\label{eq:l}
L = 2t \int \ d^2 z \ d^2 \theta \  g_{a b} (\Phi) D_{+}
\Phi^{a} D_{-} \Phi^{b}
\end{equation}

where
\begin{equation}
D_{\pm}= \frac{\del}{\del \theta^{\pm}} + i \theta^{\pm} (\frac{\del}{\del x^{0}} \pm \frac{\del}{\del x^1})
\end{equation}
\begin{equation}
\Phi^a = \phi^a + \theta^{+} \psi^a_{+} + \theta^{-} \psi^a_{-} + \theta^{-} \theta^{+} F^{a}
\end{equation}
\begin{equation}
d^2z = \frac{i}{2} dz \wedge d \bar{z}
\end{equation}

Expanding out (\ref{eq:l}) and then setting $F^{a}= \Gamma^{a}_{bc}
\psi^b_{+} \psi^c_{-}$ (the on-shell value of $F^{a}$) we have

\begin{eqnarray}
L = 2t \int d^2 z &(& \frac{1}{2} g_{a b} \del_z \phi^{a}
\del_{\bar{z}} \phi^{b} + \frac{i}{2} g_{a b} \psi^{a}_{-}
D_z \psi^{b}_{-} \nonumber\\  && {} + \frac{i}{2} g_{a b}
\psi^{a}_{+} D_{\bar{z}}  \psi^{b}_{+} + \frac{1}{4} R_{abcd} \psi^{a}_{+} \psi^{b}_{+} \psi^{c}_{-}
\psi^{d}_{-} )
\end{eqnarray}
where $D_{\bar{z}} \psi^{a}_{+} = \del_{\bar{z}} \psi^{a}_{+} +\Gamma^{a}_{b c} \  \del_{\bar{z}} \phi^{b}
\ \psi^{c}_{+}$ and $D_{z} \psi^{a}_{-} = \del_{z} \psi^{a}_{-} + \Gamma^{a}_{b c} \ \del_{z} \phi^{b} \ \psi^{c}_{-}$.

If the target space is K\"ahler the nonlinear sigma model
will have an additional $(1,1)$ supersymmetry, turning the theory into $N=(2,2)$ sigma model \cite{Zumino}.
The Lagrangian of such a sigma model is written as

\begin{eqnarray}
L = 2t \int d^2 z &(& \frac{1}{2} g_{ab} \del_z \phi^{a}
\del_{\bar{z}} \phi^{b} + i g_{\bar{\mu} \mu} \psi^{\bar{\mu}}_{-}
D_z \psi^{\mu}_{-} \nonumber\\  && {} + i g_{\bar{\mu}\mu}
\psi^{\bar{\mu}}_{+} D_{\bar{z}}  \psi^{\mu}_{+} + R_{\mu \bar{\mu} \nu \bar{\nu}} \psi^{\mu}_{+} \psi^{\bar{\mu}}_{+} \psi^{\nu}_{-}
\psi^{\bar{\nu}}_{-} )
\end{eqnarray}

The detailed supersymmetry transformations are listed as follows \cite{Witten} \footnote{This set of the transformation
laws is on-shell. The  equivalence between on-shell and off-shell formalism for topological twisted models on K\"ahler targets was first worked out by Labastida and Llatas in \cite{LLL}. It has also been shown that the off-shell formalism exists for the bi-Hermitian geometry \cite{LRvUZ}.
In the main part of this paper we will present our results in an on-shell way and collect the off-shell calculation in the appendix. }.
\begin{eqnarray}
\label{eq:longvar}
&& \delta \phi^{\mu} = i \epsilon_{-} \psi^{\mu}_{+} + i  \epsilon_{+} \psi^{\mu}_{-}  \nonumber \\
&& \delta \phi^{\bar{\mu}} = i \bar{\epsilon}_{-} \psi^{\bar{\mu}}_{+} + i  \bar{\epsilon}_{+} \psi^{\bar{\mu}}_{-}  \nonumber \\
&& \delta \psi^{\mu}_{+} = - \bar{\epsilon}_{-}  \del_z \phi^{\mu} - i \epsilon_{+} \psi^{\nu}_{-} \Gamma^{\mu}_{\nu \sigma} \psi^{\sigma}_{+}\nonumber \\
&& \delta \psi^{\bar{\mu}}_{+} = - \epsilon_{-}  \del_z \phi^{\bar{\mu}} -i\bar{\epsilon}_{+} \psi^{\bar{\nu}}_{-} \Gamma^{\bar{\mu}}_{\bar{\nu} \bar{\sigma}} \psi^{\bar{\sigma}}_{+}\nonumber \\
&& \delta \psi^{\mu}_{-} = - \bar{\epsilon}_{+}  \del_{\bar{z}} \phi^{\mu} -i \epsilon_{-} \psi^{\nu}_{+} \Gamma^{\mu}_{\nu \sigma}\psi^{\sigma}_{-}\nonumber \\
&& \delta \psi^{\bar{\mu}}_{-} = - \epsilon_{+}  \del_{\bar{z}} \phi^{\bar{\mu}} - i \bar{\epsilon}_{-} \psi^{\bar{\nu}}_{+} \Gamma^{\bar{\mu}}_{\bar{\nu} \bar{\sigma}}\psi^{\bar{\sigma}}_{-}
\end{eqnarray}

\subsection{K\"ahler A model}
An A-twist will turn $\psi^{\mu}_{+}$ and $\psi^{\bar{\mu}}_{-}$ into sections of $\Phi^{*}(T^{1,0}X)$ and
$\Phi^{*}(T^{0,1}X)$, denoted as $\chi^{\mu}$ and $\chi^{\bar{\mu}}$. And $\psi^{\bar{\mu}}_{+}$ and $\psi^{\mu}_{-}$ become sections of $\Omega^{1,0}_{\Sigma} \otimes \Phi^{*}(T^{0,1}X)$
and $\Omega^{0,1}_{\Sigma} \otimes \Phi^{*}(T^{1,0}X)$, denoted as $\psi^{\bar{\mu}}_z$ and $\psi^{\mu}_{\bar{z}}$.
In order to get the transformation laws we simply set $\epsilon_{+}=\bar{\epsilon}_{-}=0$ in (\ref{eq:longvar}). After A-twist the Lagrangian becomes

\begin{eqnarray}
L = 2t \int d^2 z &(& \frac{1}{2} g_{ab} \del_z \phi^{a}
\del_{\bar{z}} \phi^{b} +i g_{\bar{\mu} \mu} \psi^{\bar{\mu}}_{z}
D_{\bar{z}} \chi^{\mu} \nonumber\\  && {} + i g_{\bar{\mu} \mu}
\psi^{\mu}_{\bar{z}} D_{z}  \chi^{\bar{\mu}} - R_{\mu \bar{\mu} \nu\bar{\nu}} \psi^{\mu}_{\bar{z}} \psi^{\bar{\mu}}_{z} \chi^{\nu} \chi^{\bar{\nu}} )
\end{eqnarray}

The key fact as stated in \cite{Witten} is that the Lagrangian can be recast into a
very suggestive form, which is a BRST exact term plus a pullback term depdending only on the K\"ahler structure of the target space.  Upon deriving this the equatoins of motion of $\psi$ are needed.

\begin{equation}
\label{eq:cool}
L= it \int \ d^2z  \{Q, V_A  \} +t \int \Phi^{*} (K)
\end{equation}
with $V_A= g_{\mu \bar{\nu}} ( \psi^{\bar{\nu}}_z \del_{\bar{z}} \phi^{\mu} + \del_z \phi^{\bar{\nu}} \psi^{\mu}_{\bar{z}} )$ and $K= -i g_{\mu \bar{\nu}} dz^{\mu} d z^{\bar{\nu}}$. From this expression we realize that the K\"ahler A model depends only on the cohomology class of $K$. $\int \Phi^{*} (K)$ also depends on the homotopy class of the mapping $\Phi$, but in the path integral all the homotopy classes will be summed over.

\subsection{K\"ahler B model}
We also recall some basics about the K\"ahler B model which will be useful later. The B twist will turn $\psi^{\bar{\mu}}_{\pm}$ into sections of
$\Phi^{*}(T^{0,1}X)$, and $\psi^{\mu}_{+}$ and $\psi^{\mu}_{-}$ into sections of $\Omega^{1,0}_{\Sigma} \otimes \Phi^{*}(T^{0,1}X)$
and $\Omega^{0,1}_{\Sigma} \otimes \Phi^{*}(T^{0,1}X)$ respectively.
The transformation can be written as
\begin{eqnarray}
&&\delta \phi^{\mu} = 0 \nonumber \\
&&\delta \phi^{\bar{\mu}} = i \epsilon \eta^{\bar{\mu}} \nonumber\\
&&\delta \eta^{\bar{\mu}} = \delta \theta_{\mu} = 0 \nonumber \\
&&\delta \rho^{\mu} = - \epsilon d \phi^{\mu}
\end{eqnarray}
where
\begin{eqnarray}
&&\eta^{\bar{\mu}} = \psi^{\bar{\mu}}_{+} + \psi^{\bar{\mu}}_{-} \nonumber \\
&&\theta_{\mu} =g_{\mu \bar{\mu}} (\psi^{\bar{\mu}}_{+} - \psi^{\bar{\mu}}_{-}) \nonumber \\
&& \rho^{\mu} = \psi^{\mu}_{+} + \psi^{\mu}_{-}
\end{eqnarray}
After the B twisting the Lagrangian explicitly becomes

\begin{eqnarray}
L = t \int d^2 z &(& g_{ab} \del_z \phi^{a}
\del_{\bar{z}} \phi^{b} +i g_{\bar{\mu} \mu} \eta^{\bar{\mu}} (D_z \rho^{\mu}_{\bar{z}} +
D_{\bar{z}} \rho^{\mu}_{z})  \nonumber \\&& {} + i \theta_{\mu}  (D_{\bar{z}} \rho^{\mu}_{z} -D_z \rho^{\mu}_{\bar{z}} )
+ R_{\mu \bar{\mu} \nu\bar{\nu}}  \rho^{\mu}_z \rho^{\nu}_{\bar{z}} \eta^{\bar{\mu}} \theta_{\sigma} g^{\sigma \bar{\nu}} )
\end{eqnarray}

which can be reexpressed as follows.
\begin{equation}
L = it \int \{ Q, V_B \} + tW
\end{equation}
where
\begin{equation}
W= \int_{\Sigma} ( - \theta_{\mu} D \rho^{\mu} -\frac{i}{2} R_{ \mu \bar{\mu} \nu \bar{\nu}} \rho^{\mu} \wedge \rho^{\nu} \eta^{\bar{\mu}} \theta_{\sigma} g^{\sigma \bar{\nu}} )
\end{equation}
and the $D$ operator is the exterior derivative on the worldsheet $\Sigma$ by using the pullback of the Levi-Civita connection
on $M$.
The model is topological because it is independent of the complex structure of the worldsheet and the K\"ahler structure of
the target space. However the model do depend on the complex structure, which can be seen from the BRST variations of
the fields.

\section{Bi-Hermitian geometry and its topological twisted models}
\label{biH}

As stated in the introduction the most general $(2,2)$ nonlinear sigma model with $H$ is described in
\cite{GHR}, which is also known as "bi-Hermitian geometry." We will simply quote the properties of the
geometry, without any derivations of the requirements. With the non-trivial B-field turned on, the worldsheet
action is given by

\begin{equation}
\label{eq:baction} L = 2t \int \ d^2 z \ d^2 \theta \  (g_{a
b}(\Phi) + b_{ab}(\Phi) ) D_{+} \Phi^{a} D_{-} \Phi^{b}
\end{equation}

The first set of $(1,1)$ supersymmetry is as usual while
the additional $(1,1)$ supersymmetry transformations are given by
two different complex structures

\begin{eqnarray}
\label{eq:delta12}
\delta^1 \Phi^{a} = i \epsilon^1_{+} D_{+} \Phi^{a} + i \epsilon^1_{-} D_{-} \Phi^{a} \nonumber \\
\delta^2 \Phi^{a} = i \epsilon^2_{+} D_{+} \Phi^{b} J_{+b}^{a}
+ i \epsilon^2_{-} D_{-} \Phi^{b} J_{- b}^{a}
\end{eqnarray}

where $J_{+}$ and $J_{-}$ are the complex structures seen by the
left and right movers respectively. Requiring (\ref{eq:baction}) to be
invariant under the transformations leads us to the conditions:

\begin{equation}
\label{eq:J}
J_{\pm}^t g J_{\pm} = g \ \ \ \ \ \nabla^{\pm} J_{\pm} =0
\end{equation}
where $\nabla^{\pm}$ are the covariant derivatives with torsional
connections $\Gamma_{\pm} = \Gamma \pm g^{-1} H$.
The first condition implies that the metric is Hermitian with respect to the
either one of the complex structures $J_{\pm}$. And the second condition in (\ref{eq:J}) explicitly becomes
\begin{equation}
\label{eq:JJ}
J^{a}_{\pm b , c} = \Gamma^{d}_{\pm cb} J^{a}_{\pm d} - \Gamma^{a}_{\pm cd} J^{d}_{\pm b} .
\end{equation}
Equation (\ref{eq:JJ}) will be used when we try to construct the
generalized A/B models in real coordinate basis. Moreover the $H$
field is of type $(2,1)+(1,2)$ with respect to both complex
structures $J_{\pm}$. Expanding (\ref{eq:baction}) out and then
setting $F^a$ to its on-shell value we have the following worldsheet
action in component fields
\begin{eqnarray}
&F^a&= \Gamma^a_{+bc} \psi^{b}_{+} \psi^{c}_{-} = - \Gamma^a_{-bc} \psi^{b}_{-} \psi^{c}_{+}\\
\nonumber \\
&L& = 2 t \int d^2 z ( \frac{1}{2} (g_{ab} + b_{ab})
\del_z \phi^{a} \del_{\bar{z}} \phi^{b} + \frac{i}{2} g_{a
b} ( \psi^{a}_{-} \del_z \psi^{b}_{-} + \psi^{a}_{+}
\del_{\bar{z}}  \psi^{b}_{+}) \\  && {} + \frac{i}{2}
\psi^{a}_{-} \del_z \phi^{b}  \psi^{c}_{-} (\Gamma_{abc} - H_{abc}) + \frac{i}{2} \psi^{a}_{+}
\del_{\bar{z}} \phi^{b}  \psi^{c}_{+} (\Gamma_{abc} + H_{abc}) + \frac{1}{4} R_{+abcd} \psi^{a}_{+} \psi^{b}_{+} \psi^{c}_{-}
\psi^{d}_{-} ) \nonumber
\end{eqnarray}

where $R_{+abcd}$ is the curvature of the torsional connection $\Gamma^{a}_{+bc}$.

\begin{equation}
R_{\pm abcd} = R_{abcd} \pm \frac{1}{2} (\nabla_{d} H_{abc}  - \nabla_{c} H_{abd} ) + \frac{1}{4} (H^{e}_{ad}H_{ebc}- H^{e}_{ac}H_{ebd})
\end{equation}


Since the theory is of $(2,2)$ type there exist two $U(1)$
R-symmetries for the worldsheet fermions, $U(1)_V$ and $U(1)_A$
\cite{AK}. The topological A and B twists will shift
the spins of the fermions by the charges of $U(1)_V$ and $U(1)_A$
respectively. The charge assignments are worked out in \cite{AK} and \cite{CGJ}.

\begin{eqnarray}
U(1)_V : \ \ \ q_V(\bar{P}_{+} \psi_{+}) = -1 \ \ \ q_V(\bar{P}_{-} \psi_{-}) = -1 \nonumber \\
U(1)_A : \ \ \ q_A(\bar{P}_{+} \psi_{+}) = -1 \ \ \ q_V(\bar{P}_{-} \psi_{-}) = +1
\end{eqnarray}

with the following projectors defined for conveniences.

\begin{equation}
\label{eq:projector}
P_{\pm}= \frac{1}{2}(1 + i J_{\pm}), \ \ \ \bar{P}_{\pm}=
\frac{1}{2}(1 - i J_{\pm})
\end{equation}

Moreover the $U(1)$ R-symmetry used in the topological twist needs to be
non-anomalous. The anomalies are computed by Atiyah-Singer index theorem and the
conditions are
\begin{eqnarray}
U(1)_V : \ \ \ c_1(T^{1,0}_{-}) - c_1(T^{1,0}_{+}) =0 \nonumber \\
U(1)_A : \ \ \ c_1(T^{1,0}_{-}) + c_1(T^{1,0}_{+}) =0
\end{eqnarray}
Using the language of generalized complex geometry we have two commuting
twisted generalized complex structures $( \mathcal{J}_1, \mathcal{J}_2)$. $\mathcal{J}_1$
and  $\mathcal{J}_2$ are endomorphisms on $TM \oplus T^{*}M$, which square to $-1$.
Let $E_1$ and $E_2$ be the $i$-eigenbundles of $\mathcal{J}_1$
and  $\mathcal{J}_2$. The conditions can be repackaged into

\begin{eqnarray}
U(1)_V : \ \ \ c_1(E_2) =0 \nonumber \\
U(1)_A : \ \ \ c_1(E_1) =0
\end{eqnarray}

The supersymmetry transformation laws can be derived from (\ref{eq:delta12}).
\begin{eqnarray}
& &\delta^1_{+} \phi = \psi_{+} \ \ \ \ \ \delta^1_{-} \phi = \psi_{-} \ \ \ \  \delta^2_{+} \phi = J_{+} \psi_{+} \ \ \ \ \  \delta^2_{-}\phi = J_{-} \psi_{-} \nonumber\\
& &\delta^1_{+} \psi_{+} = - i \del_{z} \phi \ \ \ \delta^1_{-} \psi_{+} = F \ \ \  \delta^2_{+} \psi_{+} =  i J_{+} \del_z \phi \ \ \  \delta^2_{-}\psi_{+} = J_{-}F \nonumber\\
& &\delta^1_{+} \psi_{-} = -F \ \ \ \delta^1_{-} \psi_{-} = -i \del_{\bar{z}} \phi \ \ \  \delta^2_{+} \psi_{-} = -J_{+}F \ \ \ \delta^2_{-} \psi_{-} = i J_{-} \del_{\bar{z}} \phi
\end{eqnarray}

We can then define the linear combinations of the supersymmetry generators.
\begin{eqnarray}
Q_{+} =\frac{1}{2} ( Q^1_{+} + i Q^2_{+} )  \ \ \ \ \  \bar{Q}_{+} =\frac{1}{2} ( Q^1_{+} - i Q^2_{+} ) \nonumber \\
Q_{-} =\frac{1}{2} ( Q^1_{-} + i Q^2_{-} )  \ \ \ \ \  \bar{Q}_{-} =\frac{1}{2} ( Q^1_{-} - i Q^2_{-} )
\end{eqnarray}

We then express the on-shell variation laws in the following forms

\begin{eqnarray}
\label{eq:major}
\delta \phi^a &=& i (\epsilon_{+} (P_{+} \psi_{+})^a + \bar{\epsilon}_{+} (\bar{P}_{+} \psi_{+})^a) +
i (\epsilon_{-} (P_{-} \psi_{-})^a + \bar{\epsilon}_{-} (\bar{P}_{-} \psi_{-})^a) \nonumber \\
\delta \psi^{a}_{+} &=& - \epsilon_{+} (\bar{P}_{+} \del_{z} \phi)^a - \bar{\epsilon}_{+} (P_{+} \del_{z} \phi)^a
                  - \Gamma^a_{+bc} \delta \phi^{b} \psi^{c}_{+} \nonumber \\ &+& i H^a_{bc}
(\epsilon_{+} (P_{+} \psi_{+})^b + \bar{\epsilon}_{+} (\bar{P}_{+} \psi_{+})^b) \psi^{c}_{+}
-\frac{i}{2} (\epsilon_{+} P^a_{+d} + \bar{\epsilon}_{+} \bar{P}^a_{+d} ) H^d_{bc} \psi^b_{+} \psi^c_{+} \nonumber \\
\delta \psi^{a}_{-} &=& - \epsilon_{-} (\bar{P}_{-} \del_{z} \phi)^a - \bar{\epsilon}_{-} (P_{-} \del_{z} \phi)^a
                  - \Gamma^a_{-bc} \delta \phi^{b} \psi^{c}_{-} \nonumber \\ &+& i H^a_{bc}
(\epsilon_{-} (P_{-} \psi_{-})^b + \bar{\epsilon}_{-} (\bar{P}_{-} \psi_{-})^b) \psi^{c}_{-}
-\frac{i}{2} (\epsilon_{-} P^a_{-d} + \bar{\epsilon}_{-} \bar{P}^a_{-d} ) H^d_{bc} \psi^b_{-} \psi^c_{-}
\end{eqnarray}

where $\epsilon_{\pm}$ are the variation parameters of $Q_{\pm}$.

The BRST operators for the generalized A and B models can be taken as:

\begin{equation}
Q_A = Q_{+} + \bar{Q}_{-}, \ \ \ Q_B = \bar{Q}_{+} + \bar{Q}_{-}.
\end{equation}

Before the topological twists we have the worldsheet fermions
$P_{+}\psi_{+}$, $\bar{P}_{+}\psi_{+}$, $P_{-}\psi_{-}$, and
$\bar{P}_{-}\psi_{-}$. These fermions are sections of certain
bundles. For instance $\bar{P}_{+}\psi_{+}$ is a section of
$K^{1/2} \otimes \Phi^{*}(T^{0,1}_{+}X)$ where $K$ is the
canonical line bundle of the worldsheet (the bundle of $(1,0)$
form.) and $T^{0,1}_{+}$ is the $(0,1)$ part of the tangent bundle
with respect to $J_{+}$. After performing topological A-twist, the spins of the
fermions will be changed as follows.

\begin{eqnarray}
(P_{+}\psi_{+})^{a} &\equiv& \chi^{a} \in \Gamma ( \Phi^{*}(
T^{1,0}_{+}X ) ) \nonumber \\
(\bar{P}_{+}\psi_{+})^{a} &\equiv& \chi^{a}_z \in \Gamma (
\Omega^{(1,0)}_{\Sigma} \otimes \Phi^{*} (T^{0,1}_{+}X )) \nonumber \\
({P}_{-}\psi_{-})^{a} &\equiv& \lambda^{a}_{\bar{z}} \in \Gamma
(\Omega^{(0,1)}_{\Sigma} \otimes \Phi^{*} (T^{1,0}_{-}X )) \nonumber \\
(\bar{P}_{-}\psi_{-})^{a} &\equiv& \lambda^{a} \in \Gamma (
\Phi^{*}( T^{0,1}_{-}X ))
\end{eqnarray}

On the other hand the B-twist case can be obtained similarly. For
completeness we list the sections in the generalized B-model with
the BRST charge $Q_B = \bar{Q}_{+} + \bar{Q}_{-}$.

\begin{eqnarray}
\label{eq:bmodelfield}
(P_{+}\psi_{+})^{a} &\equiv& \chi^{a}_z \in \Gamma ( \Omega^{(1,0)}_{\Sigma} \otimes \Phi^{*}(
T^{1,0}_{+}X )) \nonumber \\
(\bar{P}_{+}\psi_{+})^{a} &\equiv& \chi^{a} \in \Gamma ( \Phi^{*} (T^{0,1}_{+}X )) \nonumber \\
(P_{-}\psi_{-})^{a} &\equiv& \lambda^{a}_{\bar{z}} \in \Gamma ( \Omega^{(0,1)}_{\Sigma} \otimes \Phi^{*}(
T^{1,0}_{-}X )) \nonumber \\
(\bar{P}_{-}\psi_{-})^{a} &\equiv& \lambda^{a}  \in
\Gamma ( \Phi^{*}( T^{0,1}_{-}X ))
\end{eqnarray}

\subsection{Generalized A model}
We will use the generalized A-model as our first explicit example. The BRST variation of the fields
can be written down by setting the variation of $\bar{Q}_{+}$ and $Q_{-}$ in (\ref{eq:major}) to be zero.

\begin{eqnarray}
\label{eq:Avar}
\{ Q_A, \phi^{a} \} = &&\chi^{a} +\lambda^{a}  \nonumber \\
\{ Q_A, \chi^{a} \} = && - i \Gamma^{a}_{+bc} \lambda^b \chi^c \nonumber \\
\{ Q_A, \lambda^{a} \} = && -i \Gamma^{a}_{-bc} \chi^b \lambda^c \nonumber \\
\{ Q_A, \chi^{a}_z \}  = &&- i \Gamma^{a}_{+bc} (\chi^b + \lambda^b) \chi^{c}_z \nonumber \\
                                    && - (\bar{P}_{+} \del_z \phi)^a + i \bar{P}^a_{+d} H^d_{bc} \chi^b \chi^c_z \nonumber \\
\{ Q_A, \lambda^{a}_{\bar{z}} \} = &&- i \Gamma^{a}_{-bc} (\chi^b + \lambda^b) \lambda^{a}_{\bar{z}} \nonumber \\
                                    && - (P_{-} \del_z \phi)^a - i P^a_{-d} H^d_{bc} \lambda^b \lambda^c_{\bar{z}}
\end{eqnarray}

After the twisting the Lagrangian becomes:

\begin{eqnarray}
\label{eq:atwistl}
&L& = 2 t \int d^2 z ( \frac{1}{2} (g_{ab} + b_{ab})
\del_z \phi^{a} \del_{\bar{z}} \phi^{b} + i g_{ab} ( \chi^{a}_z \del_{\bar{z}} \chi^{b} +
\lambda^{a}_{\bar{z}} \del_z \lambda^{b} )
 \\  && {} + i (\Gamma_{abc} - H_{abc}) \chi^{a}_z \del_{\bar{z}}  \phi^b \chi^{c}+
i (\Gamma_{abc} + H_{abc}) \lambda^{a}_{\bar{z}} \del_z  \phi^b \lambda^{c}+
R_{+abcd} \chi^{a} \chi^{b}_{z} \lambda^{c}_{\bar{z}} \lambda^{d} ) \nonumber
\end{eqnarray}

We mimic the $V_A$ operator in K\"ahler A model (\ref{eq:cool}) by virtue of the projectors.
\begin{equation}
\mathcal{V}_A = g_{ab} ( \chi^{a}_z ( P_{+} \del_{\bar{z}} \phi) ^{b} + \lambda^{a}_{\bar{z}}  ( \bar{P}_{-} \del_z \phi ) ^{b} )
\end{equation}

The BRST variations of $(P_{+} \del_{\bar{z}} \phi)^b$ and $(\bar{P}_{-} \del_z \phi)^b$ will involve the derivatives of the complex structures
and can be re-expressed in terms of $\Gamma_{\pm}$ and the projectors (\ref{eq:projector}) by using (\ref{eq:JJ}) and $J_{\pm} = -i (P_{\pm} - \bar{P}_{\pm} )$.

\begin{eqnarray}
\label{eq:projvar}
\{ Q_A , (P_{+} \del \phi)^b \} = \del \chi^{b} + (P_{+} \del \lambda )^{b} &+& \frac{1}{2} \Gamma^{d}_{+ec} (P_{+} - \bar{P}_{+} )^b_d ( \chi^c+ \lambda^c) \del \phi^{e} \nonumber\\  &-&\frac{1}{2} \Gamma^{b}_{+cd}  ( \chi^c+ \lambda^c)  (P_{+} \del \phi - \bar{P}_{+} \del \phi) ^d  \nonumber \\
\{ Q_A , (\bar{P}_{-} \del \phi)^b \} = \del \lambda^b + (\bar{P}_{-} \del \chi) ^b &-& \frac{1}{2} \Gamma^{d}_{-ec} (P_{-} - \bar{P}_{-} )^b_d ( \chi^c+ \lambda^c) \del \phi^{e} \nonumber\\  &+&\frac{1}{2} \Gamma^{b}_{-cd}  ( \chi^c+ \lambda^c)  (P_{-} \del \phi - \bar{P}_{-} \del \phi) ^d
\end{eqnarray}

Here the $\del$ operator could be either $\del_z$ or $\del_{\bar{z}}$.
Performing the BRST variations to $V$ by using (\ref{eq:Avar}) and (\ref{eq:projvar}) we obtain
\begin{eqnarray}
\label{eq:qv}
\{ Q_A , \mathcal{V}_A \} = i g_{ab} ((\bar{P}_{+} \del_z  \phi ) ^{a}  (P_{+} \del_{\bar{z}} \phi )^b +  ({P}_{-} \del_{\bar{z}} \phi ) ^{a} (\bar{P}_{-} \del_{z}  \phi )^b )
+ g_{ab} ( \chi^a_z \del_{\bar{z}} \chi^b + \lambda^a_{\bar{z}}\del_{z} \lambda^b)  \nonumber \\
+ (\Gamma_{abc} +H_{abc}) \chi_z^a \del _{\bar{z}} \phi^b \chi^c +  (\Gamma_{abc} -H_{abc}) \lambda_{\bar{z}}^a \del_z  \phi^b
\lambda^{c}
\end{eqnarray}

The curvature term will be recovered if we use the equations of motion for $\chi_z$ and $\lambda_{\bar{z}}$. To visualize that the model
only depends on one of the generalized complex structure one can use the following identities.

\begin{eqnarray}
g(P_{\pm} \cdot, \bar{P}_{\pm} \cdot) = \frac{1}{2} g ( \cdot, \cdot) + \frac{i}{2} g ( J_{\pm} \cdot, \cdot) =  \frac{1}{2} g ( \cdot, \cdot) + \frac{i}{2} \omega_{\pm} ( \cdot, \cdot) \nonumber \\
g(\bar{P}_{\pm} \cdot, P_{\pm} \cdot) = \frac{1}{2} g ( \cdot, \cdot) - \frac{i}{2} g ( J_{\pm} \cdot, \cdot)  =\frac{1}{2} g ( \cdot, \cdot) - \frac{i}{2} \omega_{\pm}( \cdot, \cdot)
 \end{eqnarray}
The scalar term in (\ref{eq:qv}) becomes
\begin{equation}
g_{ab} ( (\bar{P}_{+}\del_z \phi ) ^{a}  (P_{+} \del_{\bar{z}} \phi )^b + ({P}_{-} \del_{\bar{z}}  \phi ) ^{a} (\bar{P}_{-} \del_{z} \phi )^b ) =
2 g_{ab} \del_z  \phi^{a} \del_{\bar{z}} \phi^{b} - i \tilde{\omega}_{ab} \del_z  \phi^{a} \del_{\bar{z}} \phi^{b}
\end{equation}
where $\tilde{\omega}= \frac{1}{2}(\omega_{+} +\omega_{-})$ which appear in $\mathcal{J}_2$ in (\ref{eq:gcs}).

Comparing the twisted action (\ref{eq:atwistl}) and  (\ref{eq:qv}) we obtain the following suggestive equation, modulo the equations
of motion for $\chi_z$ and $\lambda_{\bar{z}}$.
\begin{equation}
\label{eq:aa}
L = it \int d^2z \ \{ Q_A, \mathcal{V}_A \} + t \int \Phi^{*} (-i \tilde{\omega} ) + t \int \Phi^{*}(b)
\end{equation}
Apparently the action of the generalized A model depends on one of the generalized complex structures $\mathcal{J}_2$ and the pullback of the spacetime $b$ field. The topological feature of the action will be made clear in the next section.

\subsection{Generalized B model}
The generalized B model has the field contents as listed in (\ref{eq:bmodelfield}). By projecting out $\epsilon_{\pm}$ in (\ref{eq:major}) the BRST variations for these fields are similarly obtained.

\begin{eqnarray}
\label{eq:Bvar}
\{ Q_B, \phi^{a} \} = &&\chi^{a} +\lambda^{a}  \nonumber \\
\{ Q_B, \chi^{a} \} = && - i \Gamma^{a}_{+bc} \lambda^b \chi^c \nonumber \\
\{ Q_B, \lambda^{a} \} = && -i \Gamma^{a}_{-bc} \chi^b \lambda^c \nonumber \\
\{ Q_B, \chi^{a}_z \}  = &&- i \Gamma^{a}_{+bc} (\chi^b + \lambda^b) \chi^{c}_z \nonumber \\
                                    && - (P_{+} \del_z \phi)^a + i P^a_{+d} H^d_{bc} \chi^b \chi^c_z \nonumber \\
\{ Q_B, \lambda^{a}_{\bar{z}} \} = &&- i \Gamma^{a}_{-bc} (\chi^b + \lambda^b) \lambda^{a}_{\bar{z}} \nonumber \\
                                    && - (P_{-} \del_z \phi)^a - i P^a_{-d} H^d_{bc} \lambda^b \lambda^c_{\bar{z}}
\end{eqnarray}
 with $Q_B= \bar{Q}_{+} + \bar{Q}_{-}$. Comparing (\ref{eq:Avar}) and (\ref{eq:Bvar}) we can see that the A and B
model variation laws are simply exchanged if we substitute $J_{+}$
by $-J_{+}$. In generalized B model the operator in the BRST exact
term is given by
\begin{equation}
\mathcal{V}_B=g_{ab} (\chi^a_z  (\bar{P}_{+} \del_{\bar{z}} \phi )^b + \lambda^a_{\bar{z}}  ( \bar{P}_{-} \del_z \phi)^b)
\end{equation}
The variations of $(\bar{P}_{\pm} \del \phi)^b$ are given by

\begin{eqnarray}
\label{eq:bprojvar}
\{ Q_B , (\bar{P}_{\pm} \del \phi)^b \} = (\bar{P}_{\pm} (\del \chi + \del \lambda) )^{b} &-& \frac{1}{2} \Gamma^{d}_{\pm ec} (P_{\pm} - \bar{P}_{\pm} )^b_d (
 \chi^c+ \lambda^c) \del \phi^{e} \nonumber\\
&+&\frac{1}{2} \Gamma^{b}_{-cd}  ( \chi^c+ \lambda^c)  (P_{\pm} \del \phi - \bar{P}_{\pm} \del \phi) ^d
\end{eqnarray}
Note that $\bar{P}_{+} \chi = \chi$ and $\bar{P}_{-} \lambda = \lambda$. Again the $\del$ could be either $\del_z$ or
$\del{\bar{z}}$.

The Lagrangian after the twisting is given by
\begin{eqnarray}
\label{eq:btwistl}
&L& = 2 t \int d^2 z ( \frac{1}{2} (g_{ab} + b_{ab})
\del_z \phi^{a} \del_{\bar{z}} \phi^{b} + i g_{ab} ( \chi^{a}_z \del_{\bar{z}} \chi^{b} +
\lambda^{a}_{\bar{z}} \del_z \lambda^{b} )
 \\  && {} + i (\Gamma_{abc} - H_{abc}) \chi^{a}_z \del_{\bar{z}}  \phi^b \chi^{c}+
i (\Gamma_{abc} + H_{abc}) \lambda^{a}_{\bar{z}} \del_z  \phi^b \lambda^{c}+
R_{+abcd} \chi^{a} \chi^{b}_{z} \lambda^{c} \lambda^{d}_{\bar{z}} ) \nonumber
\end{eqnarray}

In order to determine the pullback term we compute $\{Q, \mathcal{V}_B\}$.
\begin{eqnarray}
\label{eq:bqv}
\{Q_B, \mathcal{V}_B \} =i g_{ab} ( (P_{+}\del_z \phi ) ^{a}  (\bar{P}_{+} \del_{\bar{z}}\phi )^b + ({P}_{-} \del_{\bar{z}}\phi ) ^{a}  (\bar{P}_{-} \del_{z}\phi )^b )
+ g_{ab} ( \chi^a_z \del_{\bar{z}} \chi^b + \lambda^a_{\bar{z}}\del_{z} \lambda^b)  \nonumber \\
+ (\Gamma_{abc} +H_{abc}) \chi_z^a \del _{\bar{z}} \phi^b \chi^c +  (\Gamma_{abc} -H_{abc}) \lambda_{\bar{z}}^a \del_z  \phi^b
\lambda^{c}
\end{eqnarray}
In deriving this we have used the equations of motion of the fermionic fields. Note that (\ref{eq:qv}) and (\ref{eq:bqv}) are almost
the same except for the scalar kinetic terms. This will result in the different GCS dependence.  Namely,
\begin{equation}
\label{eq:bb}
L = it \int d^2z \ \{Q_B, \mathcal{V}_B\} + t \int \Phi^{*} ( i \delta \omega ) + t \int \Phi^{*}(b)
\end{equation}
where $\delta \omega = \frac{1}{2} (\omega_{+} - \omega_{-} )$
appearing in $\mathcal{J}_1$ (\ref{eq:gcs}). Contrary to the
generalized A model, the generalized B model depends on
$\mathcal{J}_1$. At first sight the results (\ref{eq:aa})
(\ref{eq:bb}) seem nice and confirm our original guess. A second
thought, however, reveals the issue that neither of $b -i
\tilde{\omega}$ and $b + i \delta \omega$ is closed. The consequence
of this is that under small coordinate reparametrization the
variation of the pullback will be nonvanishing and proportional to
$H$ \cite{Zucch3}. One way to solve this issue is to appeal to the
GCG \cite{Zucch2}. Working in generalized B model, we assume the
pure spinor $\circledS_1$ associated with TGC structure
$\mathcal{J}_1$ can be put into the following form:

\begin{eqnarray}
\circledS_1 = exp (b +\beta) \\
- \bar{\beta} = b \mp  i \omega_{\pm} - \gamma_{\pm}
\end{eqnarray}

where $d \beta = 0$ and the multiplication in the exponential is the wedge product.
A direct but lengthy computation shows, in generalized B model,

\begin{equation}
L = it \int d^2z \ \{Q_B, \mathcal{V}_B + \frac{1}{2} \gamma_{+ab} \chi^a_{z} \del_{\bar{z}} \phi^b -
\frac{1}{2} \gamma_{-ab} \lambda^a_{\bar{z}} \del_{z} \phi^b \} + t \int \Phi^{*} (\bar{\beta} )
\end{equation}

We refer the interested readers to \cite{Zucch2} for more details
about this construction. Alternatively one could simply say that
without this construction the model is topological in the sense that
the worldsheet metric is irrelevant.


\section{Conclusion and Discussion}
\label{con}
In this paper we study the topological twisted models with $H$-flux.
We explicitly expand the $N=(2,2)$ worldsheet action with bi-Hermitian target spaces and
twist the action. We found that the generalized twisted models have many similar features to the K\"ahler
twisted models. For example, the action can always be written as a sum of a BRST exact term
and some pullback terms, from which the geometric dependence of the topological models can be read off.
The generalized A/B model depends only on one of the twisted generalized complex structures  $\mathcal{J}_2$/
 $\mathcal{J}_1$.

Although it is very powerful to construct interesting examples of
topological field theories by "twisting" the spins of the fields,
some topological constraints for anomaly cancelations always come
with it. Recently people have tried to construct the topological
models for generalized geometries by using Batalin-Vilkovisky
formalism to get around this limitation\cite{Pestun}.

Another advantage of the twisted models is that it makes explicit the studying the mirror symmetry, in this case, of the
non-K\"ahler spaces. The lacking of the non-K\"ahler examples, however, is a long-standing problem along this
direction. Although the "generalized K\"ahler" examples provided in \cite{CKT} are not twisted by $H$-field, it
would still be very interesting to study the topological models for those geometries. Another interesting problem is
to generalize the usual K\"ahler quotients to obtain explicit bi-Hermitian examples. We would like to visit these problems in the future. \\

{\bf Acknowledgments:} Thanks go to Alessandro Tomasiello for very
useful conversations and Anton Kapustin for comments. WYC received
support from the DOE under contract DE-AC03-76SF00515.

\appendix
\section{Appendix: Off-shell $(2,2)$ worldsheet supersymmetry}
\subsection{K\"ahler case}
Consider a $(1,1)$ worldsheet action with auxiliary fields. The action
is given by

\begin{eqnarray}
L = 2t \int d^2 z &(& \frac{1}{2} g_{a b} \del_z \phi^{a}
\del_{\bar{z}} \phi^{b} + \frac{i}{2} g_{a b} \psi^{a}_{-} D_z
\psi^{b}_{-} \nonumber\\  && {} + \frac{i}{2} g_{a b} \psi^{a}_{+}
D_{\bar{z}}  \psi^{b}_{+} + \frac{1}{4} R_{abcd} \psi^{a}_{+}
\psi^{b}_{+} \psi^{c}_{-} \psi^{d}_{-} + \frac{1}{4} g_{a b}
\tilde{F}^a  \tilde{F}^b)
\end{eqnarray}

where $\tilde{F}^a = F^a - \Gamma^a_{bc} \psi^{b}_{+}\psi^{c}_{-}$.
When the target space is K\"ahler, we can easily rewrite the action
in holomorphic and anti-holomorphic coordinates and the $(1,1)$
theory is promoted to $(2,2)$.

The off-shell $(2,2)$ supersymmetry is as follows.
\begin{eqnarray}
\label{eq:longlongvar}
&& \delta \phi^{\mu} = i \epsilon_{-} \psi^{\mu}_{+} + i  \epsilon_{+} \psi^{\mu}_{-}  \nonumber \\
&& \delta \phi^{\bar{\mu}} = i \bar{\epsilon}_{-} \psi^{\bar{\mu}}_{+} + i  \bar{\epsilon}_{+} \psi^{\bar{\mu}}_{-}  \nonumber \\
&& \delta \psi^{\mu}_{+} = - \bar{\epsilon}_{-}  \del_z \phi^{\mu} - i \epsilon_{+} (\psi^{\nu}_{-} \Gamma^{\mu}_{\nu \sigma} \psi^{\sigma}_{+} + \tilde{F}^{\mu} )\nonumber \\
&& \delta \psi^{\bar{\mu}}_{+} = - \epsilon_{-}  \del_z \phi^{\bar{\mu}} -i\bar{\epsilon}_{+} (\psi^{\bar{\nu}}_{-} \Gamma^{\bar{\mu}}_{\bar{\nu} \bar{\sigma}} \psi^{\bar{\sigma}}_{+}+ \tilde{F}^{\bar{\mu}})\nonumber \\
&& \delta \psi^{\mu}_{-} = - \bar{\epsilon}_{+}  \del_{\bar{z}} \phi^{\mu} -i \epsilon_{-} (\psi^{\nu}_{+} \Gamma^{\mu}_{\nu \sigma}\psi^{\sigma}_{-}+ \tilde{F}^{\mu})\nonumber \\
&& \delta \psi^{\bar{\mu}}_{-} = - \epsilon_{+}  \del_{\bar{z}} \phi^{\bar{\mu}} - i \bar{\epsilon}_{-} (\psi^{\bar{\nu}}_{+} \Gamma^{\bar{\mu}}_{\bar{\nu} \bar{\sigma}}\psi^{\bar{\sigma}}_{-} + \tilde{F}^{\bar{\mu}})\nonumber \\
&& \delta \tilde{F}^{\mu} =  i \epsilon_{-} \Gamma^{\mu}_{\nu \sigma} \psi^{\nu}_{+} \tilde{F}^{\sigma} + i \bar{\epsilon}_{-} \Gamma^{\mu}_{\nu \sigma} \psi^{\nu}_{-} \tilde{F}^{\sigma} + i \epsilon_{-} (\text{EOM of } \psi^{\mu}_{+}) +  i \epsilon_{+} (\text{EOM of } \psi^{\mu}_{-}) \nonumber \\
&& \delta \tilde{F}^{\bar{\mu}} =  i \bar{\epsilon}_{+}
\Gamma^{\bar{\mu}}_{\bar{\nu} \bar{\sigma}} \psi^{\bar{\nu}}_{-}
\tilde{F}^{\bar{\sigma}} +i \epsilon_{+}
\Gamma^{\bar{\mu}}_{\bar{\nu} \bar{\sigma}} \psi^{\bar{\nu}}_{+}
\tilde{F}^{\bar{\sigma}}+  i \bar{\epsilon}_{+} (\text{EOM of }
\psi^{\bar{\mu}}_{-}) +  i \bar{\epsilon}_{-} (\text{EOM of }
\psi^{\bar{\mu}}_{+})
\end{eqnarray}

where $\tilde{F}^{\mu} = F^{\mu} - \Gamma^{\mu}_{\nu \sigma}
\psi^{\nu}_{+}\psi^{\sigma}_{-}$ and $\tilde{F}^{\bar{\mu}} =
F^{\bar{\mu}} - \Gamma^{\bar{\mu}}_{\bar{\nu} \bar{\sigma}}
\psi^{\bar{\nu}}_{+}\psi^{\bar{\sigma}}_{-}$. It is a doable but
tedious computation to show that this set of transformation laws
indeed realizes supersymmetry algebra without using equations of
motion and the Lagrangian is invariant up to a boundary term under
the transformation. Moreover, it can be shown that after A-twisting,
the transformation is indeed reduced to the off-shell transformation
for A-model in \cite{LLL}. After showing this, this is safe without
any doubt that we can integrate out the auxiliary fields and use the
on-shell supersymmetry transformations (\ref{eq:longvar}).

\subsection{bi-Hermitian case}

The off-shell Lagrangian for the bi-Hermitian sigma model is as
follows.
\begin{eqnarray}
&L& = 2 t \int d^2 z ( \frac{1}{2} (g_{ab} + b_{ab}) \del_z \phi^{a}
\del_{\bar{z}} \phi^{b} + g_{ab} \tilde{F}^a\tilde{F}^b+\frac{i}{2}
g_{a b} ( \psi^{a}_{-} \del_z \psi^{b}_{-} + \psi^{a}_{+}
\del_{\bar{z}}  \psi^{b}_{+}) \\  && {} + \frac{i}{2} \psi^{a}_{-}
\del_z \phi^{b}  \psi^{c}_{-} (\Gamma_{abc} - H_{abc}) + \frac{i}{2}
\psi^{a}_{+} \del_{\bar{z}} \phi^{b}  \psi^{c}_{+} (\Gamma_{abc} +
H_{abc}) + \frac{1}{4} R_{+abcd} \psi^{a}_{+} \psi^{b}_{+}
\psi^{c}_{-} \psi^{d}_{-} ) \nonumber
\end{eqnarray}
where $\tilde{F}^a= F^{a} -\Gamma^a_{+bc} \psi^{b}_{+} \psi^{c}_{-}
= F^{a} + \Gamma^a_{-bc} \psi^{b}_{-} \psi^{c}_{+}$.

Moreover, the off-shell $(2,2)$ transformations are given by

\begin{eqnarray}
\label{eq:wtf}
\delta \phi^a &=& i (\epsilon_{+} (P_{+} \psi_{+})^a + \bar{\epsilon}_{+} (\bar{P}_{+} \psi_{+})^a) +
i (\epsilon_{-} (P_{-} \psi_{-})^a + \bar{\epsilon}_{-} (\bar{P}_{-} \psi_{-})^a) \nonumber \\
\delta \psi^{a}_{+} &=& - \epsilon_{+} (\bar{P}_{+} \del_{z} \phi)^a - \bar{\epsilon}_{+} (P_{+} \del_{z} \phi)^a
                  - \Gamma^a_{+bc} \delta \phi^{b} \psi^{c}_{+} - i(\epsilon_{-} (P_{-} \tilde{F})^{a} +
\bar{\epsilon}_{-} (\bar{P}_{-} \tilde{F})^a)  \nonumber \\ &+& i H^a_{bc}
(\epsilon_{+} (P_{+} \psi_{+})^b + \bar{\epsilon}_{+} (\bar{P}_{+} \psi_{+})^b) \psi^{c}_{+}
-\frac{i}{2} (\epsilon_{+} P^a_{+d} + \bar{\epsilon}_{+} \bar{P}^a_{+d} ) H^d_{bc} \psi^b_{+} \psi^c_{+} \nonumber \\
\delta \psi^{a}_{-} &=& - \epsilon_{-} (\bar{P}_{-} \del_{z} \phi)^a - \bar{\epsilon}_{-} (P_{-} \del_{z} \phi)^a
                  - \Gamma^a_{-bc} \delta \phi^{b} \psi^{c}_{-}  +  i(\epsilon_{+} (P_{+} \tilde{F})^{a}
                  - \bar{\epsilon}_{+} (\bar{P}_{+} \tilde{F})^a)  \nonumber \\ &+& i H^a_{bc}
(\epsilon_{-} (P_{-} \psi_{-})^b + \bar{\epsilon}_{-} (\bar{P}_{-} \psi_{-})^b) \psi^{c}_{-}
-\frac{i}{2} (\epsilon_{-} P^a_{-d} + \bar{\epsilon}_{-} \bar{P}^a_{-d} ) H^d_{bc} \psi^b_{-} \psi^c_{-}
\end{eqnarray}
\begin{eqnarray}
\delta \tilde{F}^a = \Gamma^a_{bc} \delta \phi^{b} \tilde{F}^{c} &+&
i \epsilon_{+} ( P_{+} ( \text{EOM of } \psi_{-}))^{a} +  i \epsilon_{-} ( P_{-} (\text{EOM of } \psi_{+}))^{a} \nonumber \\
& &   i\bar{\epsilon}_{+} ( \bar{P}_{+} ( \text{EOM of } \psi_{-}))^{a} +  i \bar{\epsilon}_{-} ( \bar{P}_{-} (\text{EOM of } \psi_{+}))^{a}
\end{eqnarray}

It is a generic feature that the auxiliary fields always appear in a
Gaussian form so that we always can integrate them out. Similarly,
the supersymmetry variation for $\tilde{F}$ has terms involving
equations of motion of the fermionic fields and linear terms in
itself. After integrating out $\tilde{F}$, we recover our on-shell
result (\ref{eq:major}). This justifies the use of the on-shell
formalism, as it should.

\section{Appendix: Generalized complex geometry}
In the appendix we give a short summary of the definitions of (twisted)
generalized complex structure (GC or TGC for short). Let $M$ be an even
dimensional manifold and $H$ be a closed 3-form on $M$. The twisted
Dorfman backet $\circ$ is defined as a binary operation on the sections of $TM \oplus T^{*}M$.
\begin{equation}
(X \oplus \zeta) \circ (Y \oplus \eta) = [ X, Y ] \oplus ( \mathcal{L}_{X} \eta - \imath_Y d \zeta + \imath_Y \imath_X
H)
\end{equation}
where $X,Y \in \Gamma(TM)$ and $\zeta, \eta \in \Gamma(T^{*}M)$.
The bundle $TM \oplus T^{*}M$ has a metric $h$ with $(n,n)$ signature defined by an inner
product for the sections in $TM \oplus T^{*}M$.

A TGC-structure on $M$ is an endomorphism $\mathcal{J}$ on $TM \oplus T^{*}M$ such that

(1) $\mathcal{J}^2 = -1$

(2) $h(\cdot, \cdot) = h(\mathcal{J} \cdot , \mathcal{J} \cdot)$

(3) The $i$-eigenbundle of $\mathcal{J}$ is closed (or involutive) with respect to the twisted Dorfman
bracket. This condition is equivalent to an integrability condition for the (T)GC-structure.

Setting $H=0$ the word "twisted" is dropped everywhere and we will get the definitions for Dorfman brackets and GC-structures. \\

(Twisted) generalized K\"ahler structure consists of two commuting (T)GC-structures
$\mathcal{J}_1$ and $\mathcal{J}_2$ such that $\mathcal{G} = -\mathcal{J}_1 \mathcal{J}_2$
is a positive definite metric on $TM \oplus T^{*}M$.

A (twisted) generalized K\"ahler structure is physically relevant because it has
been shown that the structure is
equivalent to the bi-Hermitian geometry \cite{Gual}. The two
(twisted) commuting generalized complex structures $\mathcal{J}_1$
and $\mathcal{J}_2$ can be expressed in terms of the data of the
bi-Hermitian geometry, namely, $(J_{+}, J_{-}, g, H)$.

\begin{equation}
\label{eq:gcs}
\mathcal{J}_1 = \left( \begin{array}{cc} \tilde{J} & - \alpha \\
\delta \omega & - \tilde{J}^{t} \end{array} \right), \ \ \ \ \ \
\mathcal{J}_2 = \left( \begin{array}{cc} \delta J & - \beta \\
\tilde{\omega} & - \delta J^{t} \end{array} \right)
\end{equation}

where
\begin{eqnarray}
\tilde{J} = \frac{1}{2}(J_{+}+J_{-}), \ \ \ \beta =
\frac{1}{2}(\omega^{-1}_{+} + \omega^{-1}_{-}), \ \ \
\tilde{\omega} = \frac{1}{2} (\omega_{+} + \omega_{-}), \nonumber
\\
\delta J = \frac{1}{2}(J_{+}-J_{-}), \ \ \ \alpha =
\frac{1}{2}(\omega^{-1}_{+} - \omega^{-1}_{-}), \ \ \ \delta
\omega = \frac{1}{2} (\omega_{+} - \omega_{-}).
\end{eqnarray}

\begin{equation}
\omega_{\pm} (\cdot, \cdot) = g (J_{\pm} \cdot, \cdot)
\end{equation}

The $H$ is preserved by $J_{\pm}$ in the sense that the following constraints are satisfied and moreover
it is of $(2,1)+(1,2)$ type with respect to both $J_{\pm}$.
\begin{eqnarray}
H (X, Y, Z) = H ( J_{\pm} X, J_{\pm}Y , Z) +H ( J_{\pm}X, Y , J_{\pm}Z ) + H(X,J_{\pm}Y,J_{\pm}Z) \\
H(J_{\pm}X,J_{\pm}Y,J_{\pm}Z) =H(J_{\pm}X,Y,Z) +H(X,J_{\pm}Y,Z)  +H(X,Y,J_{\pm}Z)
\end{eqnarray}
The following identity is useful in deriving equations.
\begin{equation}
H( X, Y, Z) = \mp d \omega_{\pm} ( J_{\pm} X, J_{\pm} Y, J_{\pm} Z)
\end{equation}

\end{document}